\documentstyle[11pt,epsfig]{article}

\renewcommand{\theequation}{\arabic{equation}}
\newcommand{\bee}{\begin{equation}}
\newcommand{\ene}{\end{equation}}
\newcommand{\bea}{\begin{array}}
\newcommand{\ena}{\end{array}}
\newcommand{\beqa}{\begin{eqnarray}}
\newcommand{\enqa}{\end{eqnarray}}
\newcommand{\bean}{\begin{eqnarray*}}
\newcommand{\eean}{\end{eqnarray*}}

\newcommand{\dy} {\displaystyle}

\def\ru1{\rule[-0.4truecm]{0mm}{1truecm}}
\def\lo{{\scriptscriptstyle LO}}
\def\nlo{{\scriptscriptstyle NLO}}
\def\ns{{\scriptscriptstyle NS}}
\def\als{{\alpha_s}}
\def\alp{\frac{\als(Q^2)}{\pi}}
\def\q0{Q_0^2}
\def\rapplo{\left(\frac{\als^{\scriptscriptstyle LO}
(Q^2)}{\als^{\scriptscriptstyle LO} (\q0)}\right)} 
\def\rappnlo{\left(\frac{\als^{\scriptscriptstyle NLO}
(Q^2)}{\als^{\scriptscriptstyle NLO} (\q0)}\right)} 
\def\upa{\uparrow}

\def\A{{\bf A}}
\def\B{{\bf B}}
\def\C{{\bf C}}
\def\D{{\bf D}}
\def\E{{\bf E}}
\def\F{{\bf F}}
\def\G{{\bf G}}
\def\al{\alpha}
\def\be{\beta}
\def\ga{\gamma}
\def\ep{\epsilon}
\def\lam{\lambda}
\def\th{\Theta}
\def\a{^{\ast}}

\def\up#1{\leavevmode \raise.16ex\hbox{#1}}

\def\sqr#1#2{{\vcenter{\vbox{\hrule height.#2pt
	\hbox{\vrule width.#2pt height#1pt \kern#1pt
	  \vrule width.#2pt}
	\hrule height.#2pt}}}}

\newcommand{\jou}[4]{{\rm #1 }{\bf #2} \up(19#3\up) #4}

\newcounter{appendix}
\newcommand{\appendice}
{
\setcounter{equation}{0}
\renewcommand{\theequation}{A.\arabic{equation}}
\addtocounter{appendix}{1}
\vspace{.7truecm}\noindent
{\Large\bf Appendix}
\vspace{.5truecm}
\noindent
}

\def\thebibliography#1{{\bf REFERENCES\markboth
 {REFERENCES}{REFERENCES}}\list
 {[\arabic{enumi}]}{\settowidth\labelwidth{[#1]}\leftmargin\labelwidth
 \advance\leftmargin\labelsep
 \usecounter{enumi}}
 \def\newblock{\hskip .11em plus .33em minus -.07em}
 \sloppy
 \sfcode`\.=1000\relax}

\begin{document}

\title{\hfill $\mbox{\small{\begin{tabular}{r}
${\rm \textstyle Bari-TH/97-292}$\\
${\rm \textstyle CPT-97/P.3578}$\\
${\rm\textstyle DSF-T-58/97}$\\
${\rm\textstyle hep-ph/9803229}$
\end{tabular}
}}$ \\[1truecm]
POLARIZED QUARKS, GLUONS AND SEA IN NUCLEON STRUCTURE FUNCTIONS}

\author{C. BOURRELY$^\dag$, F. BUCCELLA$^{\ast}$, O. PISANTI$^{\ast}$, P.
SANTORELLI$^{\diamond\dag}$, \\[.3truecm]
and J. SOFFER$^\dag$} 
\date{$~$}

\maketitle

\thispagestyle{empty}

\begin{center}
\begin{tabular}{l}
$^{\dag}$
Centre de Physique Th\'eorique, 
CNRS - Luminy, Case 907, \\
~~F-13288 Marseille Cedex 9 - France. \\
$^{\ast}$  Dipartimento di Scienze Fisiche, Universit\`a di Napoli, 
Mostra d'Oltremare, Pad.19,\\
~~I-80125, Napoli, Italy; INFN, Sezione di Napoli, Napoli, Italy. \\
$^{\diamond}$ Dipartimento di Fisica, Universit\`a di Bari, 
Via G. Amendola, 173, I-70126 Bari, Italy; \\
~~INFN, Sezione di Bari, Bari, Italy.
\end{tabular}
\end{center}

\begin{abstract}
We perform a NLO analysis of polarized deep inelastic scattering data to
test two different solutions to the so called {\it spin crisis}: one of
them based on the axial gluon anomaly and consistent with the Bjorken sum
rule and another one, where the defects in the spin sum rules and in the
Gottfried sum rule are related. In this case a defect is also expected for
the Bjorken sum rule. The first solution is slightly favoured by the SLAC
E154 results, but both options seem to be consistent with the CERN SMC data.
\end{abstract}

\bigskip

{\bf PACS} numbers: 13.60.-r, 13.60.Hb, 13.88.+e, 14.20.Dh

\newpage

\section{Introduction}

The EMC experiment on $g_1^p(x)$ \cite{emc} was giving for the first
moment of the spin dependent proton structure function $g_1^p(x)$ at
$<\!\!Q^2\!\!> = 10\,GeV^2$
\bee
\Gamma_1^p = \int_0^1 g_1^p(x)~ dx = 0.126\pm 0.010\pm 0.015,
\label{e:emc}
\ene
a value much smaller than the prediction of the Ellis-Jaffe sum rule
\cite{elljaf} for the proton (we use $a_3 = G_A/G_V = 1.2573\pm 0.0028$ and
$a_8 = 3\,F-D = 0.579\pm 0.025$ \cite{fd}), 
\bee
\Gamma_{1\,EJ}^p = \frac{F}{2} - \frac{D}{18} = 0.185\pm 0.003.
\label{e:ej}
\ene
This has stimulated a considerable experimental activity in measuring the
polarized structure functions (SF) $g_1^p(x)$, $g_1^n(x)$ ($g_1^{^3H\!e}$)
and $g_1^d(x)$ \cite{slac,smc,herm}. 

From the theoretical point of view, the hypothesis has been formulated
\cite{anom} that a relevant isoscalar contribution, related, in the
chiral-invariant factorization scheme \cite{cheng}, to a large gluon
polarization, contributes to Eq.~(\ref{e:ej}) at $O(\als)$, such as
\bee
\Gamma_1^p (Q^2) = \frac{1}{12} \left( a_3 + \frac{a_8}{3} + \frac{4}{3}
a_0 (Q^2) \right) \left( 1-\frac{\als(Q^2)}{\pi} \right) - \frac{\als
(Q^2)}{6\, \pi} \Delta G (Q^2)~.
\ene
This would explain the defect in the Ellis-Jaffe sum rule for the proton,
without affecting the more theoretically founded Bjorken sum rule
\cite{bjork}, 
\bee
\left( \Gamma_1^p - \Gamma_1^n \right)_{LO} = \frac{a_3}{6} = \frac{1}{6}
\frac{G_A}{G_V}. 
\label{e:bjorlo}
\ene
QCD corrections play an important role in modifying the r.h.s. of
Eq.~(\ref{e:bjorlo}) which is, strictly speaking, valid in the limit
$Q^2\rightarrow 0$, and reads 
\bee
(\Gamma_1^p - \Gamma_1^n) (Q^2) = \frac{1}{6} \frac{G_A}{G_V} \left( 1 -
\alp \right), 
\label{e:bjor}
\ene
where only the $O(\als)$ has been retained. The validity of the Bjorken sum
rule is so universally accepted that, in the analysis of the experimental
data on polarized SF \cite{nlofits1,alfor,nlofits2,fitjacpo}, it is almost
always assumed, rather than tested. There are few exceptions to this
attitude, as for example the paper by Altarelli {\it et al.} \cite{alfor},
where $a_3$ is evaluated to be $1.19\pm 0.09$, and the l.h.s. of
Eq.~(\ref{e:bjor}) at $Q^2 = 4\,GeV^2$ is $0.177\pm 0.014$, in agreement
with the predicted value at $O(\als)$, $0.189\pm 0.002$.

An alternative interpretation of the defect shown by Eq.~(\ref{e:emc}) has
been given \cite{pauli} by relating the defects in the Ellis-Jaffe sum rule
for the proton and the Gottfried sum rule \cite{gottfr} for the isovector
unpolarized SF of the nucleons, 
\bee
\int_0^1 \frac{F_2^p(x) - F_2^n(x)}{x}~ dx = \frac{1}{3} + \frac{2}{3}~ 
\int_0^1 [\bar u(x)-\bar d(x)]~ dx = \frac{1}{3},
\ene
to be compared with the measurement performed by the NMC experiment
\cite{nmc}, namely $0.235\pm 0.026$. The defect in the Gottfried sum rule
may be ascribed to a flavour asymmetry in the sea of the proton, which does
not mean the breaking of the isospin symmetry since the proton is not an
isoscalar. A flavour asymmetry of the parton sea in the proton was
advocated many years ago \cite{feyfie}, in connection with the Pauli
principle and an essential asymmetry in the proton which contains two
valence $u$ and one valence $d$ quark. The sign of the asymmetry is just
the right one to account for the defect in the Gottfried sum rule, $\bar
d-\bar u>0$, while to reproduce the value given by NMC one should have
\bee
\bar d-\bar u = \int_0^1 [\bar d(x)-\bar u(x)]~ dx = 0.15\pm 0.04.
\label{e:udna51}
\ene
The conservation of the SU(3) non-singlet weak currents gives
\bee
\Delta u_{val} = 2\, F, \quad\quad \Delta d_{val} = F-D.
\ene
Since in the proton $u^\upa$ is the valence parton with the highest first
moment (the quark model sum rules would imply $u^\upa = 1+F = 1.459\pm
0.006$), it is reasonable to assume, inspired by the Pauli principle and
Eq.~(\ref{e:udna51}), that it is just this parton which receives less
contribution from the sea and to write 
\bee
\Delta u_{val} = 2 F + \bar u - \bar d.
\ene
This implies a defect of about 0.033 (0.008) in the Ellis-Jaffe sum rule 
for the proton (neutron) and, what would be very relevant, a defect of 
about 0.025 in the Bjorken sum rule. In this framework one should expect 
\bee
\bea{rcl}
\Gamma_1^p &=& \dy \frac{F}{2} - \frac{D}{18} + \frac{2}{9} (\bar u-\bar d)
= 0.152\pm 0.010, \vspace{.2truecm} \\ 
\Gamma_1^n &=& \dy \frac{F}{3} - \frac{2 D}{9} + \frac{1}{18} (\bar u-\bar
d) = - 0.033\pm 0.004, 
\ena
\ene
in reasonable agreement with experiment.

Our strategy, to make a test for these two different interpretations, will
be to compare with experimental data the predictions of parton
distributions, whose properties will be dictated by these two assumptions.
Since the data extend on a wide $Q^2$ range, we need a method to solve the
integro-differential evolution equations \cite{dglap} at the
next-to-leading order (NLO). To this aim, we choose to reconstruct the SF
by using a truncated Jacobi polynomials expansion \cite{jacobi1,jacobi2}.
The main advantage of this method is that it is fast, since one can
analytically calculate the moments (Mellin transforms) of the SF which
enter the expansion. Indeed, it was already used in Ref. \cite{fitjacun}
for a NLO fit of the unpolarized distributions and in Ref. \cite{fitjacpo}
of the polarized ones.

The paper is organized as follows. In the next section we will present the
parameterization used to describe the SLAC data, with the different options
inspired by the two interpretations previously considered. Then we will
present the Jacobi reconstruction method and recall the QCD evolution of
the moments. Finally, we will report the results of our analysis for the
SLAC data and the QCD evolution to the SMC higher $Q^2$ value, before giving
our conclusions.

\section{Description of the SLAC data}

With respect to our previous paper \cite{evfit}, where the analysis of the
data was performed at leading order in $\als$ and in the chiral-invariant
factorization scheme, here we take advantage of the knowledge of the NLO
anomalous dimensions and coefficient functions within the $\overline{MS}$
renormalization scheme, which is the most usual one chosen in the
gauge-invariant factorization scheme. Therefore, instead of Eq.~(13) of our
previous paper, we write (we assume a flavour symmetric sea, $\Delta
u_{sea} = \Delta \bar u_{sea} = \Delta d_{sea} = \Delta \bar d_{sea} =
\Delta s_{sea} = \Delta \bar s_{sea} \equiv \Delta q_{sea}$):
\bee
\bea{rcl}
x \Delta u_{val} (x, \q0) &=& \eta_u A_u x^{a_v} (1-x)^{b_u} (1+\ga_v x),
\vspace{.2truecm} \\ 
x \Delta d_{val} (x, \q0) &=& \eta_d A_d x^{a_v} (1-x)^{b_d} (1+\ga_v x),
\vspace{.2truecm} \\ 
x \Delta q_{sea} (x, \q0) &=& x \Delta \bar q_{sea} (x, \q0) = \eta_s A_s
x^{a_s} (1-x)^{b_s} (1+\ga_s x), \vspace{.2truecm} \\ 
x \Delta G (x, \q0)   &=& \eta_G A_G x^{a_G} (1-x)^{b_G} (1+\ga_G x), 
\ena
\label{e:distr}
\ene
at $\q0 = 4\,GeV^2$, where $\eta_q$ ($q\,=\,u,\,d,\,s,\,G$) are the first
moments of the distributions and $A_q = A_q(a_q,b_q,\ga_q$) is given by 
\bee
A_q^{-1} =\int_0^1 dx x^{a_q - 1} (1-x)^{b_q} (1+\ga_q x) = \left( 1 +
\ga_q \frac{a_q}{a_q+b_q+1}\right) \frac{\Gamma(a_q)
\Gamma(b_q+1)}{\Gamma(a_q+b_q+1)}, 
\ene
in such a way that
\bee
\int_0^1 dx A_q x^{a_q-1} (1-x)^{b_q} (1+\ga_q x) = 1.
\ene
We fix $\ga_s=\ga_G=0$, since it is not possible to fit the values of these
parameters with sufficient accuracy.

We fit the asymmetries 
\bee
A_1^{p(n)} (x,Q^2) = \frac{g_1^{p(n)} (x,Q^2)}{F_1^{p(n)} (x,Q^2)} =
\frac{g_1^{p(n)} (x,Q^2)}{F_2^{p(n)} (x,Q^2)} 2x(1+R^{p(n)} (x,Q^2)),
\label{e:a1}
\ene
which are just the quantities experimentally measured. The choice of $A_1$
would also allow one to minimize the higher twist contribution (that should
partly cancel in the ratio in Eq.~(\ref{e:a1})), instead of cutting the
data at low $Q^2$. In constructing the ratio in Eq.~(\ref{e:a1}), we use
the MRS parameterization at $\q0=4\, GeV^2$ \cite{mrs} for the unpolarized
distributions.
The deuteron asymmetry, $A_1^d (x,Q^2)$, is given by
\bee
A_1^d (x,Q^2) = \frac{g_1^d (x,Q^2)}{F_1^d (x,Q^2)},
\ene
with
\bee
g_1^d (x,Q^2) = \frac{1}{2} \left( 1 - \frac{3}{2} \omega_D \right)
(g_1^p(x,Q^2) + g_1^n(x,Q^2)),
\ene
where $\omega_D = 0.058$ \cite{omD} takes into account the small D-wave
component in the deuteron ground state. We only include in the fit the data
obtained by the SLAC experiments \cite{slac}. The HERMES data \cite{herm}
are not used since their lower precision does not change the results. We
will compare the results with the SMC proton and deuteron data \cite{smc},
by evolving the distributions to the $Q^2$ of the CERN experiment.

In all the cases we will fix
\bee
\eta_d = F - D = -0.339\pm 0.013,
\ene
as expected from the conservation of the SU(3) non singlet weak currents.
The two different interpretations mentioned above will be characterized,
for the first option (see below, fits $\A$, $\B$, and $\C$), by the
additional constraint 
\bee
\eta_u = 2 F = 0.918\pm 0.013,
\ene
in such a way to obey the Bjorken sum rule with the first order correction,
Eq.~(\ref{e:bjor}), consistently with the NLO approximation. For the second
option (see below, fits $\D$, $\E$, $\F$, and $\G$) $\eta_u$ will be a free
parameter and the Bjorken sum rule will departure from the theoretical
expectation. The sea and gluon parameters, $\eta_s$ and $\eta_G$, will be
left free in both approaches (we impose $\eta_G\leq 3$ since, for high
values of this parameter, the $\chi^2$ shows a weak dependence on it) and
we will also consider cases with vanishing $\eta_s$ or (and) $\eta_G$ to
evaluate from the data how much their contributions is needed in both
options. We also consider a case (fit $\A$) with $\eta_s$ fixed by imposing
that the NLO value of $\Gamma_1^p$ at $<\!\!Q^2\!\!> = 3\,GeV^2$ should be
the one reported by SLAC E143, namely $0.127\pm 0.011$. Further limitations
on the parameters are:
\bee
\bea{c}
a_v > a_s > a_G \vspace{.2truecm} \\ 
b_u >3.96, ~~~~~b_d > 4.41, ~~~~~b_s > 10.1, ~~~~~b_G > 6.06,
\ena
\ene
in order to satisfy the positivity constraints with respect to the
unpolarized distributions. 

\section{The Jacobi reconstruction method}

There are different methods to solve the evolution equations \cite{dglap}
for the distributions (or the SF). One possibility is to numerically solve
them in the $x$-space \cite{numev1,numev2}. A faster choice is based on the
use of the Mellin moments,
\bee
F_n(Q^2) \equiv \int_0^1 x^{n-1}~ F(x,Q^2)~ dx,
\ene
since they are analytically calculable for distributions like the ones in
Eq.~(\ref{e:distr}). After transforming the integro-differential equations
to the $n$-space, they become ordinary differential equations in the $Q^2$
variable, which are analytically solved, and the solution is then
numerically inverted to give the distributions in the $x$-space. 

An alternative choice, which avoids the numerical inversion, is to express
the SF by means of orthogonal polynomials \cite{orthpol}. On the one hand,
this method is more favorable because it allows one to use the analytical
expressions of the Mellin moments; on the other hand, a positive feature is
also their weak dependence on the values of the reconstructed function
outside the region where data are collected.

The expansion of the SF in terms of Jacobi polynomials,
$\th_k^{\al\be}(x)$, takes the following form: 
\bee
F(x,Q^2) = x^\al (1-x)^\be~ \sum_{k=0}^\infty~ a_k^{(\al\be)} (Q^2)
\th_k^{\al\be}(x) 
\label{e:exp1}
\ene
where the Jacobi polynomials satisfy the orthogonality condition
\bee
\int_0^1~ x^\al (1-x)^\be~ \th_k^{\al\be} \th_l^{\al\be}~ dx = \delta_{kl},
\ene
and the coefficients $a_k^{(\al\be)}(Q^2)$ are defined by
\bee
a_k^{(\al\be)}(Q^2) = \int_0^1~ F(x,Q^2)~ \th_k^{\al\be} (x)~ dx.
\label{e:coef}
\ene
The expansion in Eq.~(\ref{e:exp1}) becomes useful if we consider the power
expansion of Jacobi polynomials, 
\bee
\th_k^{\al\be}(x) = \sum_{j=0}^k~ c_j^{(k)} (\al,\be)~ x^j,
\ene
with ($(\al)_k\equiv \Gamma (\al+k)/\Gamma (\al)$ is the Pochhammer symbol)
\bee
c_j^{(k)} (\al,\be) = (-1)^j \left( 
	\bea{c}
	k \\
	j
	\ena
\right) \frac{(\al+1)_k (\al+\be+k+1)_j}{(\al+1)_j} \sqrt{\frac{(\al+\be+2
k+1)\, \Gamma (\al+\be+k+1)}{\Gamma (\al+k+1)\, \Gamma (\be+k+1) k!}}.
\label{e:cjac}
\ene
By inserting it in Eq.~(\ref{e:coef}), we get
\bee
a_k^{(\al\be)} (Q^2) = \sum_{j=0}^k~ c_j^{(k)} (\al,\be)~ F_{j+1} (Q^2),
\ene
and obtain for $F(x,Q^2)$
\bee
F(x,Q^2) = x^\al (1-x)^\be~ \sum_{k=0}^\infty~ \th_k^{\al\be}(x)~
\sum_{j=0}^k~ c_j^{(k)} (\al,\be)~ F_{j+1} (Q^2). 
\label{e:fjac}
\ene
This formula is very useful because the $Q^2$ dependence of $F(x,Q^2)$ is
contained in its moments, for which the solution of the evolution equations
up to the NLO is well known. Evidently, one has to approximate
Eq.~(\ref{e:fjac}) by truncating the infinite series to a finite number of
terms, 
\bee
F^{(N)}(x) = x^\al (1-x)^\be~ \sum_{k=0}^N~ J_k^{\al\be}(x)~ \sum_{j=0}^k~
c_j^{(k)} (\al,\be)~ F_{j+1}. 
\ene
This truncation implies that outside the range where data have their values
the approximation can fail, and it is possible that oscillations take
place; however, by choosing a suitable value of $N$, one can reconstruct
the function with sufficient precision. In our analysis, we reconstructed
the unpolarized SF with $N=16$, $\al=-0.99$, and $\be=4.03$, while for the
polarized ones we used $N=8$ and allowed $\al$ and $\be$ to vary. The
values found in the different cases considered below are reported in Table
\ref{t:albe}. 

\section{Evolution of the QCD moments}
\label{s:evol}

Given the Mellin moments of the charge-conjugation even (odd) non-singlet,
singlet and gluon unpolarized distribution, at $\q0$,
\bee
\bea{rcl}
Q_{\ns\, n}^\eta (\q0) &\equiv& \dy \int_0^1~ x^{n-1} Q_\ns^\eta (x,\q0)~
dx, \quad\quad (\eta=\pm 1) \vspace{.2truecm} \\ 
\Sigma_n (\q0) &\equiv& \dy \int_0^1~ x^{n-1} \Sigma(x,\q0)~ dx,
\vspace{.2truecm} \\ 
G_n (\q0) &\equiv& \dy \int_0^1~ x^{n-1} G(x,\q0)~ dx,
\ena
\label{e:momun}
\ene
their evolution with $Q^2$ is given by the renormalization group equations.
In the $\overline{MS}$ scheme one gets, for example for the non-singlet 
distribution,
\bee
Q_{\ns\, n}^\eta (Q^2) \!=\! Q_{\ns\, n}^\eta (\q0)
\rappnlo^\frac{\ga_\ns^{(0) n}}{2\, \be_0} \left[ \!1 + \frac{\als^\lo
(Q^2) \!-\! \als^\lo (\q0)}{4\, \pi} \left( \frac{\ga_\ns^{(1) n\eta}}{2\,
\be_0} - \frac{\ga_\ns^{(0) n} \be_1}{2\, \be_0^2} \right) \right].
\ene
The complete expressions for the evolution of the distributions can be
found in the Appendix. $\ga_\ns^{(i) n}$ and $\ga_{\psi\psi}^{(i) n}$ (the
last one appears in Eq.~(\ref{e:momunev})) enter the unpolarized anomalous
dimensions for the non-singlet and singlet operators at NLO, in the
following perturbative expansion,
\bee
\ga_a^{n\eta} = \frac{\als}{4\, \pi} \left( \ga_a^{(0) n} + \frac{\als}{4\,
\pi} \ga_a^{(1) n\eta} \right),
\label{e:andim}
\ene
($\eta=\pm 1$ and $\eta=1$ for the non-singlet and the singlet
respectively), whose expressions was calculated in
\cite{andimun1,andimun2,andimun3}. We also have
\bee
\bea{rcl}
\dy \frac{\als^\nlo (Q^2)}{4\, \pi} &=& \dy \frac{1}{\be_0\,
ln\frac{Q^2}{\Lambda^2}} - \frac{\be_1}{\be_0^3}\frac{ln ln
\frac{Q^2}{\Lambda^2}}{ln^2 \frac{Q^2}{\Lambda^2}}, \vspace{.2truecm} \\
\be_0 = 11 - \frac{2}{3} n_f, &\quad\quad& \be_1 = 102 - \frac{38}{3} n_f, 
\ena
\label{e:alphas}
\ene
which is the QCD coupling constant at NLO (the LO expression is obtained
taking $\be_1=0$ in Eq.~(\ref{e:alphas})). According to the number of active
flavors $n_f$, the value of $\Lambda_\nlo^{(n_f)}$ is modified. We use
$\Lambda_\nlo^{(5)} = 0.2263$, so to have $\als (M_Z^2) = 0.118$
\cite{als}. From the matching of $\als$ at the quark thresholds we get, for
$m_b = 4.5\,GeV$, $\Lambda_\nlo^{(4)} = 0.326$ and $\Lambda_\nlo^{(3)} =
0.3797$, for $m_c = 1.5\,GeV$.

Analogous expressions to Eqs.~(\ref{e:momun}) and (\ref{e:momunev}) can be 
written for the polarized distributions, where the polarized anomalous 
dimensions can be found in \cite{nlofits1}. The polarized non-singlet and
singlet distributions are defined as follows:
\bee
\bea{rcl}
Q_\ns^p &\equiv& \dy a_3 + \frac{a_8}{3} = \frac{4}{3} \Delta u_{val} -
\frac{2}{3} \Delta d_{val}, \vspace{.2truecm} \\ 
Q_\ns^n &\equiv& \dy - a_3 + \frac{a_8}{3} = - \frac{2}{3} \Delta u_{val} +
\frac{4}{3} \Delta d_{val}, \vspace{.2truecm} \\ 
\Sigma &\equiv& a_0 = \Delta u_{val} + \Delta d_{val} + 6 \Delta q_{sea}, 
\ena
\ene
For the evolution of the polarized moments we used the fixed-flavour
scheme \cite{fixed,nlofits1} with $n_f=3$.

Since the coefficients $c_j^{(k)}$ in Eq.~(\ref{e:cjac}) contain in the
denominator higher and higher factorials as $N$ gets large, we have to take
into account the corrections to the anomalous dimensions described in Ref.
\cite{jacobi2}. They depend on the type of distribution we consider, even
or odd ($+$ or $-$ prescription, respectively), and consist in the
substitutions in the anomalous dimensions reported in the Appendix.

\section{Data analysis}

By considering the SLAC data of the E142, E143, and E154 experiments
\cite{slac}, we get the results reported in Tables \ref{t:fit1} and
\ref{t:fit2}. 

A tentative study of Tables \ref{t:fit1} and \ref{t:fit2} is very
instructive. It shows that one is able to get satisfactory fits to the data
with assumptions inspired from both the two different options we are
considering, namely $\eta_u$ and $\eta_d$ fixed to the values expected by
SU(3) non singlet current conservation, with the Bjorken sum rule obeyed
and $\eta_s$ given by the defect in the EJ sum rule (fit $\A$), or only
$\eta_d$ fixed to that value and without sea and gluon contribution for the
option inspired by the role of the Pauli principle (fit $\D$). The first
option is more successful in describing the neutron ($^3H\!e$) data (a
total contribution to the $\chi^2$ of 7 instead of 20 for the 19 data
points of E142 and E154) and deuteron data (a contribution to the $\chi^2$
of 55 instead of 63 for 56 data points), while the second one is slightly
preferable for the proton data (a contribution to the $\chi^2$ of 55
instead of 74 for 63 data points). We also note that, with the first
option, by taking $\eta_s=0$ we get a rather bad fit which is fit $\C$. 

\begin{figure}[t]
\begin{center}
\epsfig{file=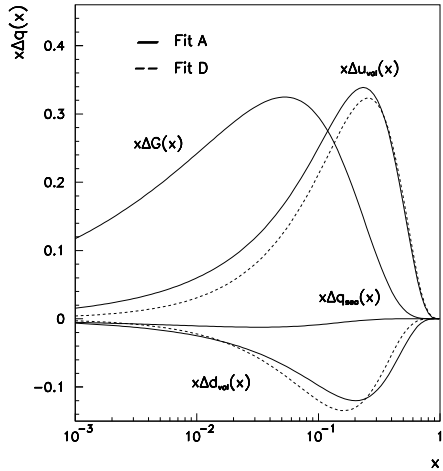,height=6.5truecm}\quad
\end{center}
\caption{\small The polarized parton distributions for fits $\A$ and $\D$
at $\q0=4\, GeV^2$. Note that, in the case of fit $\D$, sea and gluons are
vanishing.} 
\label{f:dist}
\end{figure}

\begin{figure}[t]
\begin{center}
\epsfig{file=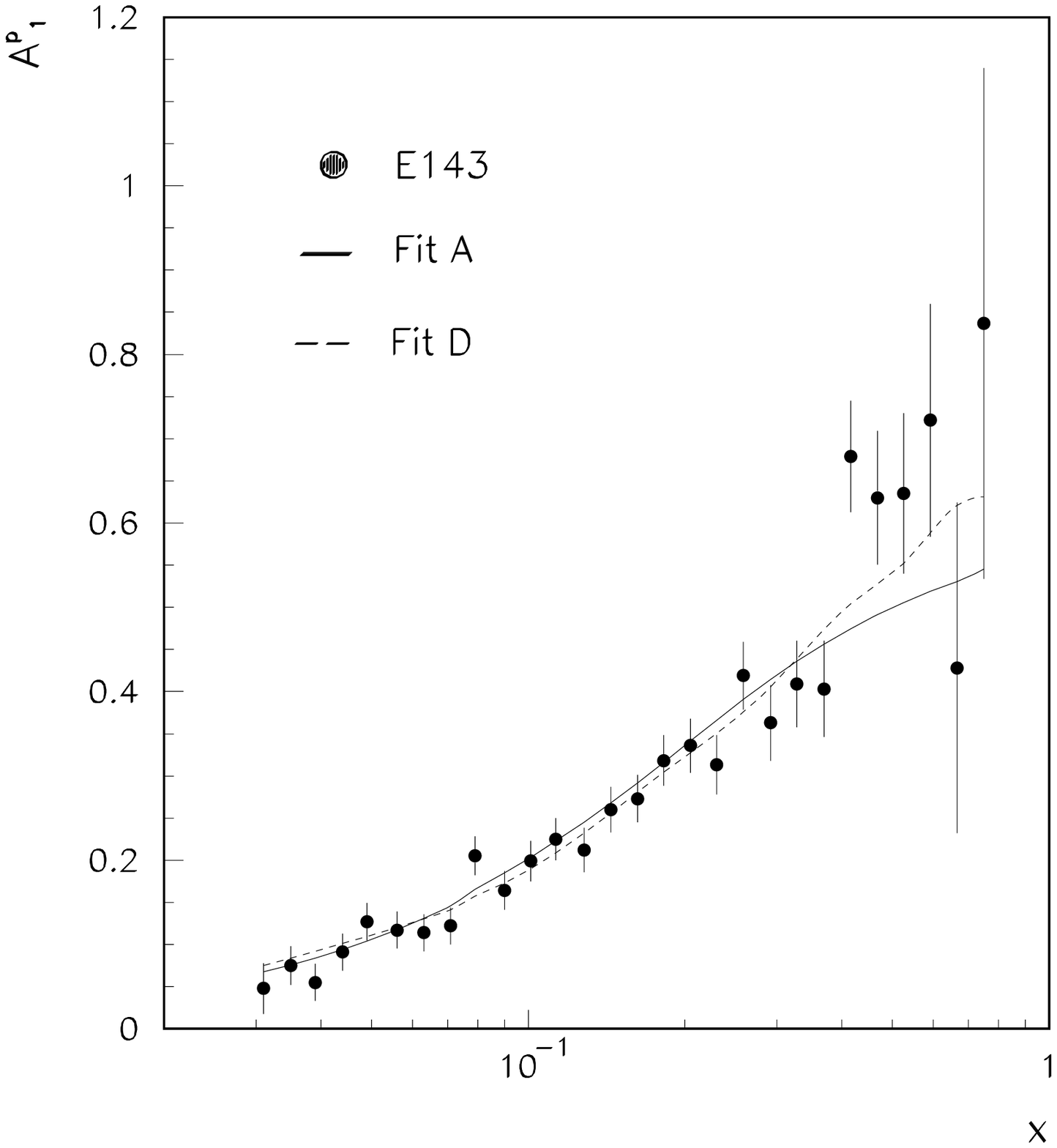,height=5.5truecm}\quad
\epsfig{file=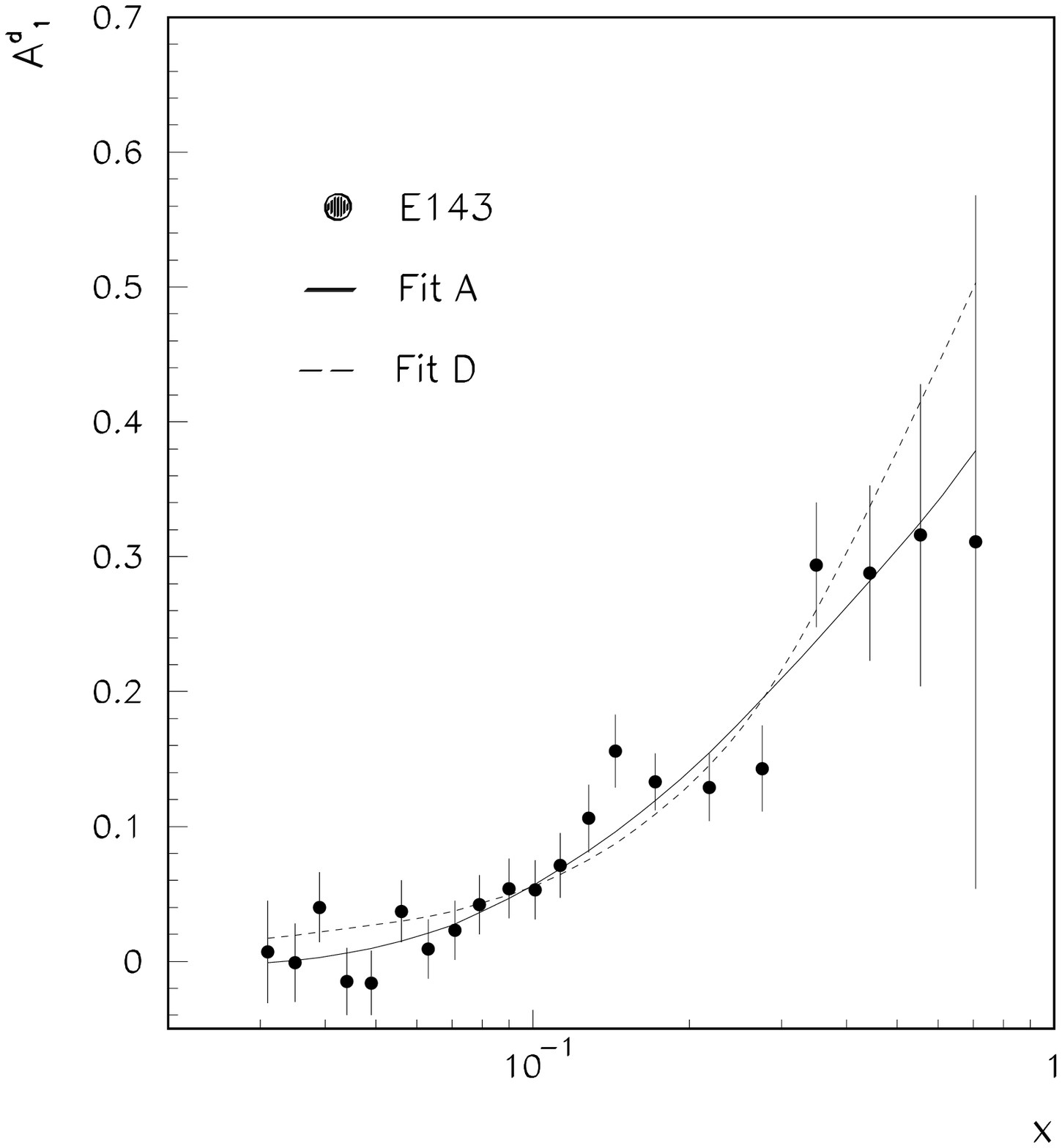,height=5.5truecm}\quad
\epsfig{file=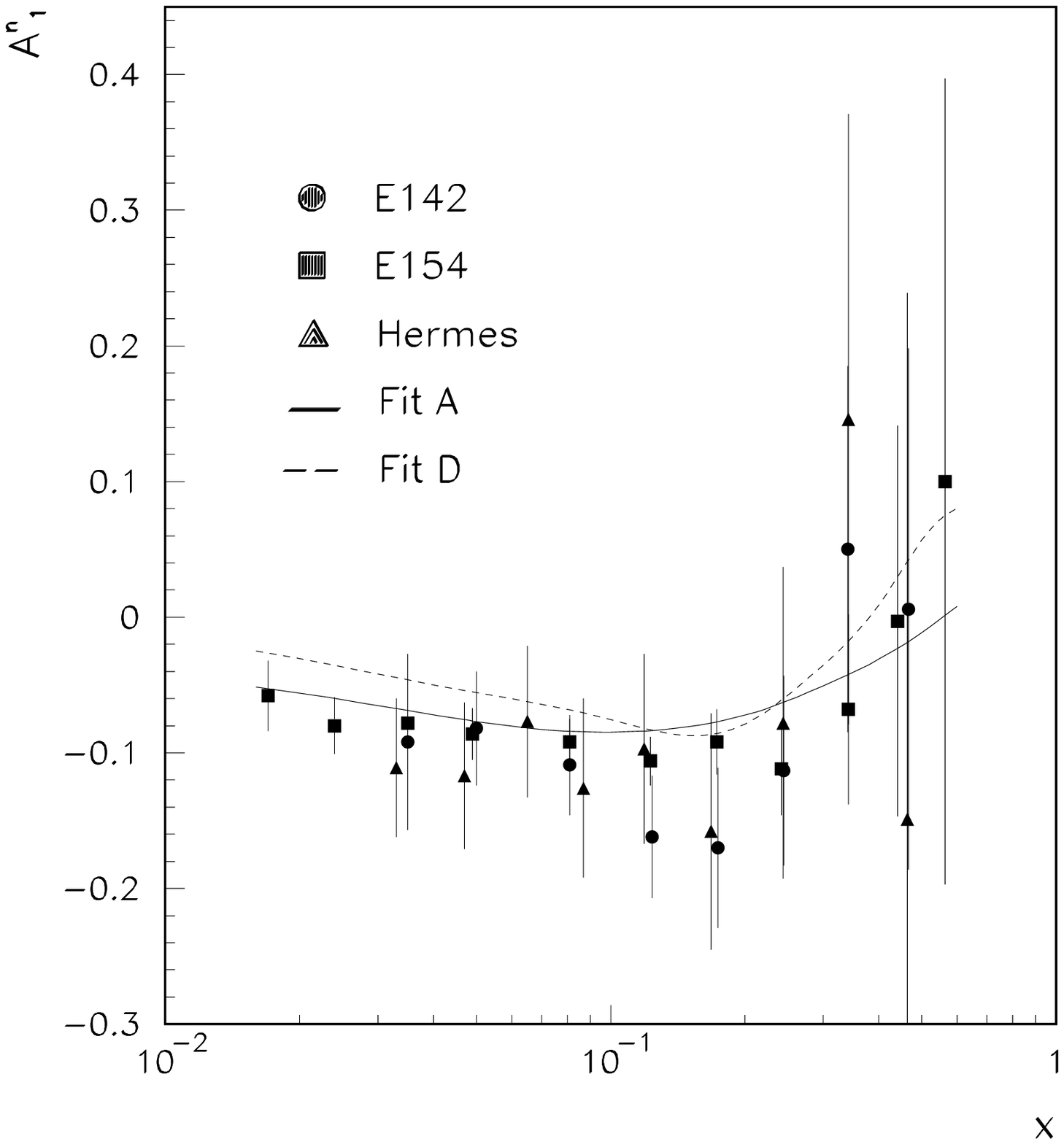,height=5.5truecm}
\end{center}
\caption{\small The prediction of fits $\A$ and $\D$ with data are showed
for proton, deuteron, and neutron asymmetries. The curves correspond to the
fits at the $Q^2$ of the data points for proton and deuteron, and at
$\q0=4\, GeV^2$ for neutron. The HERMES data \protect\cite{herm} are
showed, but they are not used in the fits.}
\label{f:fits}
\end{figure}

Fits $\A$ and $\D$ differ in the $x$ dependence of the valence parton
distribution which is more singular at $x\rightarrow 0$ ($x^{-0.427}$) for
fit $\A$ than for fit $\D$ ($x^{-0.119}$). They also differ in the
behaviour for $x\rightarrow 1$, where one has a faster power decreasing of
the $d$ partons for fit $\D$ ($(1-x)^{6.50}$) than for fit $\A$
($(1-x)^{4.66}$). It is appealing, from the point of view of the
interpretation inspired by the Pauli principle, to find in the case of the
fit $\D$, $\Delta u_{val} = 0.741 \simeq 0.918 - 0.150 = 0.768$, as
expected by the connection to the defect in the Gottfried sum rule, but it
is worth to remark that the better description of the $^3H\!e$ data by fit
$\A$ seems more relevant than the better description of the proton by fit
$\D$. When we leave free $\eta_u$ and $\eta_s$, with or without gluons
(fits $\F$ and $\G$), we find slightly better fits, with an almost average
verdict for $\eta_u$ in fit $\F$: 
\bee
\bea{rcl}
\eta_u^\F &=& \dy 0.822 \simeq \frac{\eta_u^\A+\eta_u^\D}{2} = 0.830,
\vspace{.2truecm} \\ 
\eta_s^\F &=& \dy -0.040 \simeq \frac{3}{5}~ \eta_s^\A, \vspace{.2truecm}
\\ 
\eta_G^\F &=& \dy 1.35 \simeq \eta_G^\A.
\ena
\ene

\begin{figure}[t]
\begin{center}
\epsfig{file=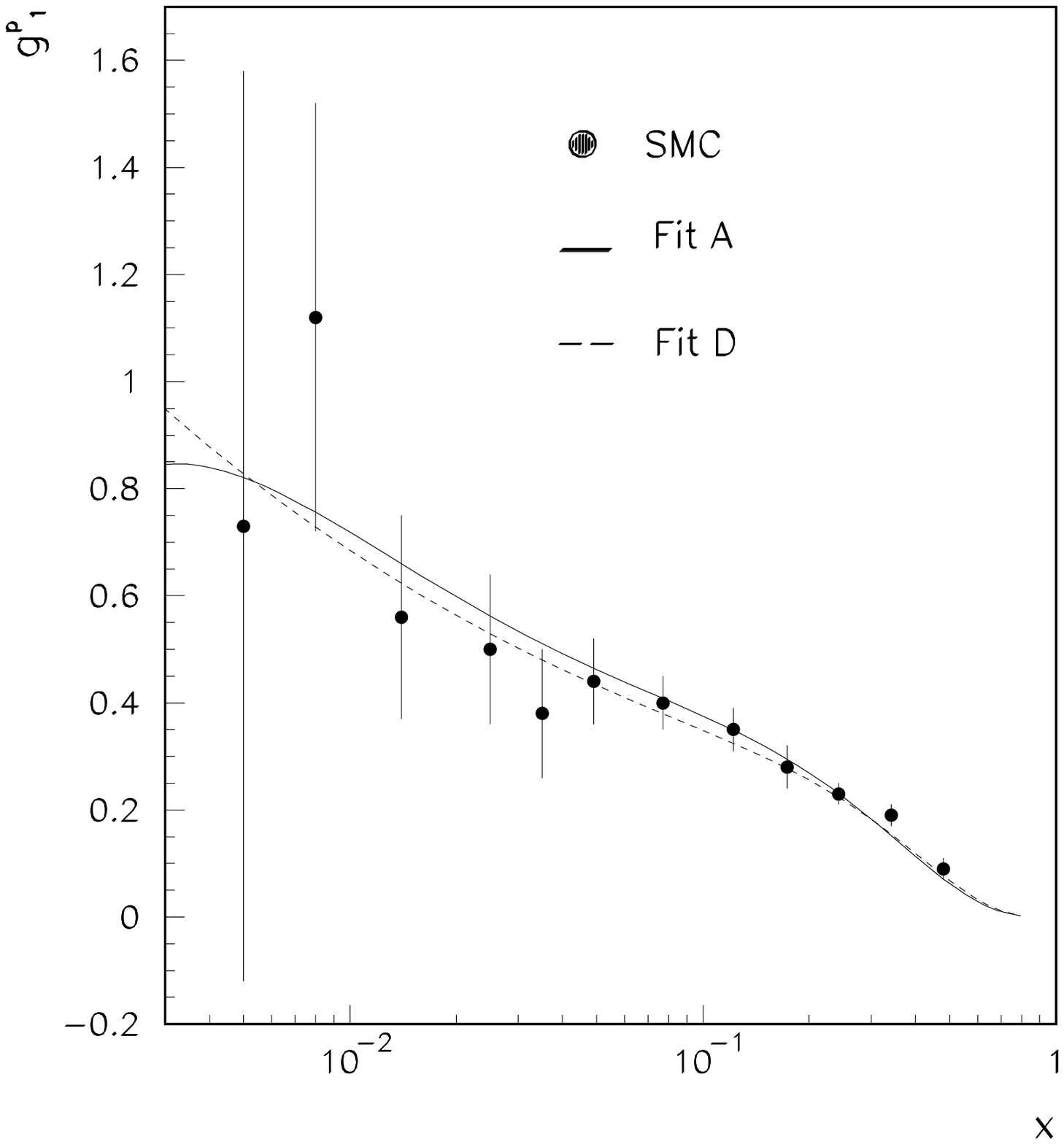,height=5.5truecm}\quad
\epsfig{file=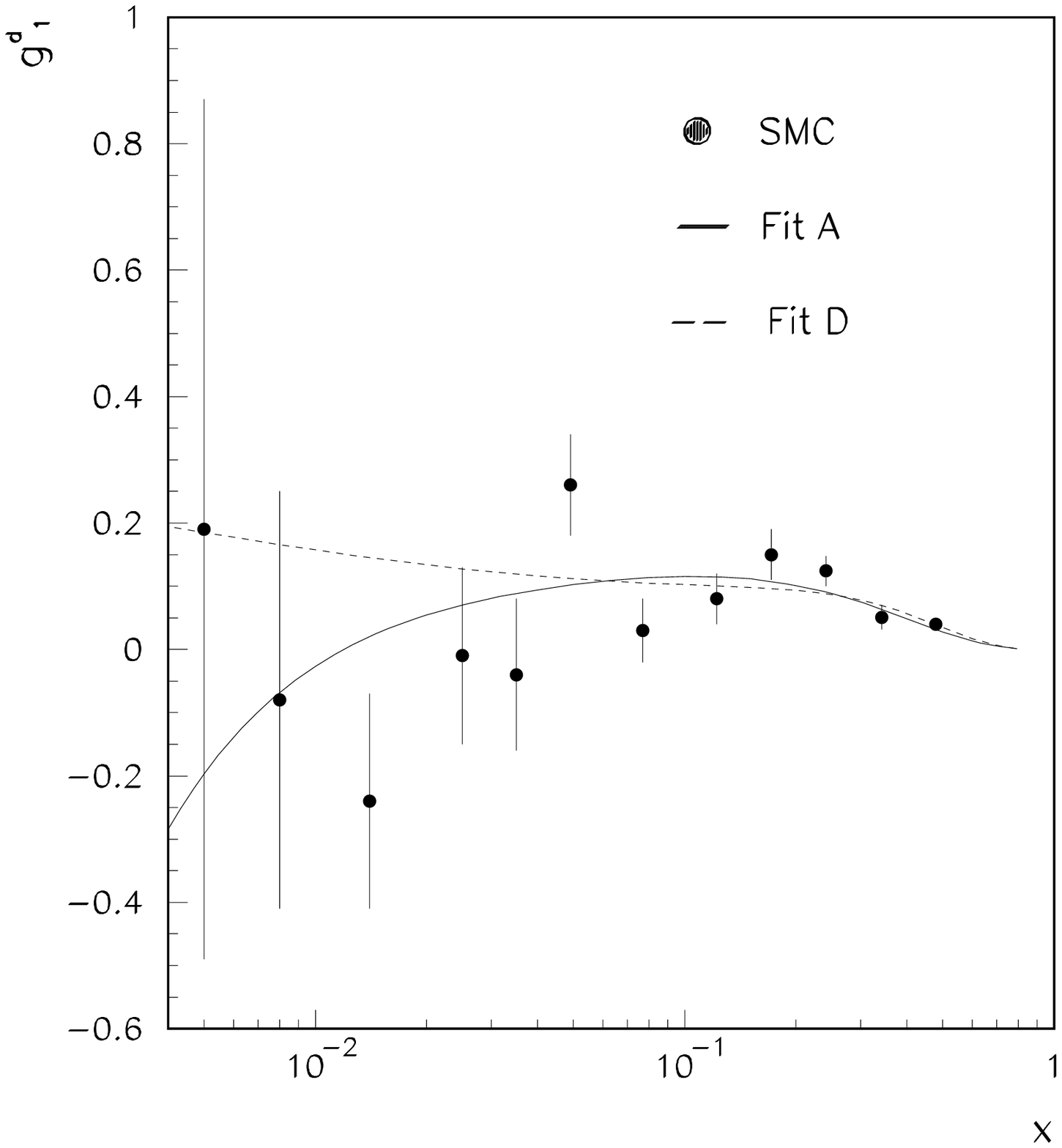,height=5.5truecm}
\end{center}
\caption{\small The predictions of fits $\A$ and $\D$ for the proton and
deuteron $g_1$ structure functions at $<\!\!Q^2\!\!> = 10\,GeV^2$ are
compared with the most recent data obtained by SMC (see Ref.
\protect\cite{smc}).}
\label{f:ev10}
\end{figure}

In Fig. \ref{f:dist} the polarized parton distributions are plotted for
fits $\A$ and $\D$ and here we note that fit $\A$ is very similar to a
solution obtained in Ref. \cite{fitjacpo} (see their Fig. 5). In Fig.
\ref{f:fits} we compare the predictions of the same fits with the SLAC
data. In Fig. \ref{f:ev10} we show the evolution of fits $\A$ and $\D$ to
$<\!\!Q^2\!\!> = 10\,GeV^2$, compared with the SMC data, which are
consistent with both of them. For this evolution we used the fortran code
by D. Fasching\footnote{We thank the author for providing us his code.}
\cite{numev1}, because the SMC data extend on a different range of $x$ and
$Q^2$, and the previously calculated $\al$ and $\be$ in the Jacobi
reconstruction of the structure functions would not allow one to well
describe them.

\section{Conclusions}

We have studied the most accurate available data on polarized deep
inelastic scattering with the purpose of testing two possible theoretical
interpretations of the {\it spin crisis}. As a result of our analysis, we
can conclude that the interpretation related to the Pauli principle is not
contradicted by the rather precise SLAC data which, however, slightly
favour the generally accepted solution with the Bjorken sum rule obeyed and
a sea contribution to the proton spin, in the gauge-invariant factorization
scheme. Obviously, a better determination of the small $x$ region could
help to clarify further the present situation.

\newpage

\appendice

The complete solutions of the evolution equations for the non-singlet,
singlet and gluon distributions, in the $\overline{MS}$ scheme, are:

\bee
\bea{rcl}
Q_{\ns\, n}^\eta (Q^2) &=& \dy Q_{\ns\, n}^\eta (\q0)
\rappnlo^\frac{\ga_\ns^{(0) n}}{2\, \be_0} \left[ 1 + \frac{\als^\lo (Q^2)
- \als^\lo (\q0)}{4\, \pi} \left( \frac{\ga_\ns^{(1) n\eta}}{2\, \be_0} -
\frac{\ga_\ns^{(0) n} \be_1}{2\, \be_0^2} \right) \right],
\vspace{.14truecm} \\
\Sigma_n (Q^2) &=& \dy \left[ (1-\al_n)~ \Sigma_n (\q0) - \tilde\al_n~ G_n
(\q0) \right] \rappnlo^\frac{\lam_+^n}{2\, \be_0} \cdot 
\vspace{.14truecm} \\ 
&& \!\!\!\!\!\!\!\!\!\!\!\!\!\!\!\!\!\!\!\!\!\!\!
\dy \cdot \left\{ 1 + \frac{\als^\lo (Q^2) - \als^\lo (\q0)}{4\, \pi}
\left( \frac{\ga_{++}^{(1) n}}{2\, \be_0} - \frac{\lam_+^n \be_1}{2\,
\be_0^2} \right) + \left[ \frac{\als^\lo (\q0)}{4\, \pi}
\rapplo^\frac{\lam_-^n-\lam_+^n}{2\, \be_0} + \right. \right.
\vspace{.14truecm} \\ 
&& \dy \left. \left. - \frac{\als^\lo (Q^2)}{4\, \pi} \right]
\frac{\ga_{+-}^{(1) n}}{2\, \be_0 + \lam_+^n - \lam_-^n} \right\} +
\vspace{.14truecm} \\
&+& \dy \left[ \al_n~ \Sigma_n (\q0) - \tilde\al_n~ G_n (\q0) \right]
\rappnlo^\frac{\lam_-^n}{2\, \be_0} \cdot \vspace{.14truecm} \\ 
&& \!\!\!\!\!\!\!\!\!\!\!\!\!\!\!\!\!\!\!\!\!\!\!
\dy \cdot \left\{ 1 + \frac{\als^\lo (Q^2) - \als^\lo (\q0)}{4\, \pi}
\left( \frac{\ga_{--}^{(1) n}}{2\, \be_0} - \frac{\lam_-^n \be_1}{2\,
\be_0^2} \right) + \left[ \frac{\als^\lo (\q0)}{4\, \pi}
\rapplo^\frac{\lam_+^n-\lam_-^n}{2\, \be_0} + \right. \right.
\vspace{.14truecm} \\ 
&& \dy \left. \left. - \frac{\als^\lo (Q^2)}{4\, \pi} \right]
\frac{\ga_{-+}^{(1) n}}{2\, \be_0 + \lam_-^n - \lam_+^n} \right\},
\vspace{.14truecm} \\ 
G_n (Q^2) &=& \dy \left[ \al_n~ G_n (\q0) - \ep_n~ \Sigma_n (\q0) \right]
\rappnlo^\frac{\lam_+^n}{2\, \be_0} \cdot \vspace{.14truecm} \\ 
&& \!\!\!\!\!\!\!\!\!\!\!\!\!\!\!\!\!\!\!\!\!\!\!
\dy \cdot \left\{ 1 + \frac{\als^\lo (Q^2) - \als^\lo (\q0)}{4\, \pi}
\left( \frac{\ga_{++}^{(1) n}}{2\, \be_0} - \frac{\lam_+^n \be_1}{2\,
\be_0^2} \right) + \left[ \frac{\als^\lo (\q0)}{4\, \pi}
\rapplo^\frac{\lam_-^n-\lam_+^n}{2\, \be_0} + \right. \right.
\vspace{.14truecm} \\ 
&& \dy \left. \left. - \frac{\als^\lo (Q^2)}{4\, \pi} \right]
\frac{\ga_{+-}^{(1) n}}{2\, \be_0 + \lam_+^n - \lam_-^n}
\frac{\ga_{\psi\psi}^{(0) n} - \lam_-^n}{\ga_{\psi\psi}^{(0) n} - \lam_+^n}
\right\} + \vspace{.14truecm} \\ 
&+& \dy \left[ (1-\al_n)~ G_n (\q0) - \ep_n~ \Sigma_n (\q0) \right]
\rappnlo^\frac{\lam_-^n}{2\, \be_0} \cdot \vspace{.14truecm} \\ 
&& \!\!\!\!\!\!\!\!\!\!\!\!\!\!\!\!\!\!\!\!\!\!\!
\dy \cdot \left\{ 1 + \frac{\als^\lo (Q^2) - \als^\lo (\q0)}{4\, \pi}
\left( \frac{\ga_{--}^{(1) n}}{2\, \be_0} - \frac{\lam_-^n \be_1}{2\,
\be_0^2} \right) + \left[ \frac{\als^\lo (\q0)}{4\, \pi}
\rapplo^\frac{\lam_+^n-\lam_-^n}{2\, \be_0} + \right. \right.
\ena
\label{e:momunev}
\ene

\[
\bea{rcl}
&& \dy \left. \left. - \frac{\als^\lo (Q^2)}{4\, \pi} \right]
\frac{\ga_{-+}^{(1) n}}{2\, \be_0 + \lam_-^n - \lam_+^n}
\frac{\ga_{\psi\psi}^{(0) n} - \lam_+^n}{\ga_{\psi\psi}^{(0) n} - \lam_-^n}
\right\}.
\ena
\]
The quantities $\al_n$, $\tilde\al_n$, $\ep_n$, $\lam_\pm^n$,
$\ga_{\pm\pm}$,  $\ga_\ns^{(i) n}$ and $\ga_{\psi\psi}^{(i) n}$  are
defined and given in \cite{buras}. The corrections to the anomalous
dimensions referred in Section \ref{s:evol} consist in the following
substitutions \cite{corr}:
\bee
\bea{rcl}
(-1)^n &\rightarrow& \pm 1 \vspace{.2truecm} \\
S_2'(\frac{n}{2}) &\rightarrow& (-1)^n \left\{ \pm S_2'(\frac{n}{2}) +
\eta_\pm (n) \left[-2 S_2(n) + \zeta(2) \right] \right\}, \vspace{.2truecm}
\\ 
S_3'(\frac{n}{2}) &\rightarrow& (-1)^n \left\{ \pm S_3'(\frac{n}{2}) +
\eta_\pm (n) \left[-4 S_3(n) + 3 \zeta(3) \right] \right\},
\vspace{.2truecm} \\ 
\tilde S(n) &\rightarrow& (-1)^n \left[ \pm \tilde S(n) + \eta_\pm (n)
\frac{5}{8} \zeta(3) \right],
\ena
\ene
where the series $S_k(n)$, $S_k'(n/2)$, and $\tilde S(n)$ are defined for
example in Ref. \cite{andimun2} and $\zeta$ is the Riemann zeta function.

\newpage

\begin{table}[ht]
\begin{center}
TABLE \ref{t:albe} \\
\vspace{.6truecm}
\begin{tabular}{||c||c|c||} 
\hline\hline\ru1
     & $\al$ & $\be$ \\
\hline\hline\ru1
\A & $0.173$ & $3.03$ \\
\hline\ru1
\B & $1.14$ & $4.20$ \\
\hline\ru1
\C & $0.265$ & $2.27$ \\
\hline\ru1
\D & $1.37$ & $4.32$ \\
\hline\ru1
\E & $1.34$ & $4.21$ \\
\hline\ru1
\F & $0.247$ & $3.45$ \\
\hline\ru1
\G & $1.36$ & $0.157$ \\
\hline\hline
\end{tabular}
\end{center}
\caption{\small The values of the parameters $\al$ and $\be$, defined in
Eq.~(\protect\ref{e:exp1}), for the polarized structure functions,
corresponding to the different fits (see section 3) are reported.}
\label{t:albe}
\end{table}

\begin{table}[ht]
\begin{center}
TABLE \ref{t:fit1} \\
\vspace{.6truecm}
\begin{tabular}{||c||c|c|c||} 
\hline\hline\ru1
     &  \A & \B & \C \\
\hline\hline\ru1
$\eta_u$ & $0.918\pm 0.013\a$ & $0.918\pm 0.013\a$ & $0.918\pm 0.013\a$ \\
\hline\ru1
$\eta_d$ & $-0.339\pm 0.013\a$ & $-0.339\pm 0.013\a$ & $-0.339\pm 0.013\a$
\\
\hline\ru1
$\eta_s$ & $-0.065\pm 0.019\a$ & $-0.050\pm 0.015$ & $0\a$ \\
\hline\ru1
$\eta_G$ & $1.69\pm 0.64$ & $0\a$ & $3.00\pm 0.08$ \\
\hline\ru1
$a_v$ & $0.573\pm 0.057$ & $0.587\pm 0.057$ & $0.320\pm 0.046$ \\ 
\hline\ru1
$a_s$ & $0.338\pm 0.080$ & $0.587\pm 0.063$ & - \\
\hline\ru1
$a_G$ & $0.338\pm 0.082$ & - & $0.312\pm 0.106$ \\
\hline\ru1
$b_u$ & $3.96\pm 0.02$ & $3.96\pm 0.02$ & $3.960\pm 0.008$ \\
\hline\ru1
$b_d$ & $4.66\pm 0.35$ & $4.71\pm 0.35$ & $5.79\pm 0.37$ \\
\hline\ru1
$b_s$ & $10.1\pm 0.8$ & $10.1\pm 0.3$ & - \\
\hline\ru1
$b_G$ & $6.06\pm 0.22$ & - & $8.54\pm 1.08$ \\
\hline\ru1
$\ga_v$ & $7.53\pm 2.06$ & $7.38\pm 1.93$ & $28.3\pm 7.7$ \\
\hline\ru1
$EJ_p$ & $0.128\pm 0.012$ & $0.137\pm 0.009$ & $0.167\pm 0.003$ \\
\hline\ru1
$EJ_n$ & $-0.061\pm 0.012$ & $-0.052\pm 0.009$ & $-0.022\pm 0.003$ \\
\hline\ru1
$Bj$ & $0.189\pm 0.003\a$ & $0.189\pm 0.003\a$ & $0.189\pm 0.003\a$ \\
\hline\ru1
$\chi^2/NDF$ & $1.07$ & $1.07$ & $1.64$ \\
\hline\hline
\end{tabular}
\end{center}
\caption{\small The values of the parameters found with the
parameterization in Eq.~(\protect\ref{e:distr}) for the different fits
corresponding to the first option (see text for more information) are
reported. The results for the r.h.s. of Ellis-Jaffe and Bjorken sum rules
are also showed. The asterisk $\ast$ indicates fixed values.} 
\label{t:fit1}
\end{table}

\begin{table}[ht]
\begin{center}
TABLE \ref{t:fit2} \\
\vspace{.6truecm}
\begin{tabular}{||c||c|c|c|c||} 
\hline\hline\ru1
     & \D & \E & \F & \G \\
\hline\hline\ru1
$\eta_u$ & $0.741\pm 0.016$ & $0.754\pm 0.021$ & $0.822\pm 0.024$ &
$0.820\pm 0.038$ \\
\hline\ru1
$\eta_d$ & $-0.339\pm 0.013\a$ & $-0.339\pm 0.013\a$ & $-0.339\pm 0.013\a$
& $-0.339\pm 0.013\a$ \\
\hline\ru1
$\eta_s$ & $0\a$ & $0\a$ & $-0.040\pm 0.020$ & $-0.028\pm 0.011$ \\
\hline\ru1
$\eta_G$ & $0\a$ & $3.00\pm 3.00$ & $1.35\pm 0.23$ & $0\a$ \\
\hline\ru1
$a_v$ & $0.881\pm 0.087$ & $0.853\pm 0.091$ & $0.730\pm 0.075$ & $0.682\pm
0.090$ \\ 
\hline\ru1
$a_s$ & - & - & $0.730\pm 0.163$ & $0.522\pm 0.220$ \\
\hline\ru1
$a_G$ & - & $0.053\pm 0.152$ & $0.730\pm 0.185$ & - \\
\hline\ru1
$b_u$ & $3.960\pm 0.003$ & $3.96\pm 0.06$ & $3.96\pm 0.04$ & $3.960\pm
0.003$ \\ 
\hline\ru1
$b_d$ & $6.50\pm 0.46$ & $6.25\pm 0.47$ & $5.36\pm 0.44$ & $5.41\pm 0.50$
\\ 
\hline\ru1
$b_s$ & - & - & $41.5\pm 11.1$ & $10.1\pm 0.4$ \\
\hline\ru1
$b_G$ & - & $6.06\pm 3.91$ & $6.06\pm 2.48$ & - \\
\hline\ru1
$\ga_v$ & $3.70\pm 1.56$ & $3.97\pm 1.66$ & $5.64\pm 1.89$ & $6.65\pm 2.50$
\\ 
\hline\ru1
$EJ_p$ & $0.132\pm 0.003$ & $0.134\pm 0.004$ & $0.124\pm 0.013$ &
$0.131\pm 0.010$ \\
\hline\ru1
$EJ_n$ & $-0.031\pm 0.003$ & $-0.030\pm 0.003$ & $-0.051\pm 0.012$ &
$-0.044\pm 0.007$ \\
\hline\ru1
$Bj$ & $0.162\pm 0.003$ & $0.164\pm 0.004$ & $0.175\pm 0.004$ & $0.174\pm
0.006$ \\
\hline\ru1
$\chi^2/NDF$ & $1.05$ & $1.07$ & $0.97$ & $1.01$ \\
\hline\hline
\end{tabular}
\end{center}
\caption{\small Same as in Table \protect\ref{t:fit1} for the fits
corresponding to the second option.} 
\label{t:fit2}
\end{table}

\end{document}